\documentclass{iau} 
\usepackage{graphicx}

\title[Rotation on the AGB scenario for multiple populations of Globular Clusters] 
{What Young Massive Clusters in the Magellanic Clouds teach us about Old Galactic Globular Clusters?}

\author[F. D'Antona et al.]   
{Francesca D'Antona$^1$,
 Paolo Ventura$^1$, Aaron Dotter$^2$, Sylvia Ekstr\"om$^3$ \and Marco Tailo$^4$}

\affiliation{$^1$INAF-OAR, \\ via di Frascati 33,
I-00078 Monteporzio Catone, Italy \\ email: {\tt franca.dantona@gmail.com} \\[\affilskip]
$^2$Harvard-Smithsonian Center for Astrophysics, \\60 Garden Street, Cambridge, MA 02138, USA \\email: {\tt aaron.dotter@gmail.com}\\[\affilskip]
$^3$Geneva Observatory, University of Geneva, \\Maillettes 51, CH-1290 Sauverny, Switzerland \\email:{sylvia.ekstrom@unige.ch }\\[\affilskip]
$^4$Dipartimento di Fisica e Astronomia `Galileo Galilei', Univ. di Padova, \\Vicolo dell'Osservatorio 3, I-35122 Padova, Italy  \\email: {\tt mrctailo@gmail.com}}

\pubyear{2018}
\volume{343}  
\jname{Why Galaxies care about AGB stars: a continuing challenge through cosmic time}
\editors{ F. Kerschbaum, M. Groenewegen, \& H. Olofsson, eds.}
\begin{document}

\maketitle

\begin{abstract}
The Asymptotic Giant Branch (AGB) scenario ascribes the multiple populations in old Galactic Globular Clusters (GGC) to episodes of star formation in the gas contaminated by the ejecta of massive AGBs and super-AGBs of a first stellar population. The mass of these AGBs (4-8M$_\odot$) today populate the Young Massive Clusters (YMC) of the Magellanic Clouds, where rapid rotation and its slowing down play an important role in shaping the color magnitude diagram features. 
Consequently, we must reconsider whether the rotational evolution of these masses affects the yields, and whether the resulting abundances are compatible with the chemical patterns observed in GGC. We show the first results of a differential analysis, by computing the hot bottom burning evolution of non rotating models with increased CNO-Na abundances at the 2DU, following the results of MESA rotational models.
\end{abstract}

\firstsection 
\section{Introduction}

\cite{mackey-nielsen2007} discovered that the intermediate age ($\sim$1.5\,Gyr) Globular Cluster (GC) NGC\,1846 in the Large Magellanic Cloud (LMC) displayed a spread turnoff, signalling the presence of multiple populations, possibly similar to those revealed by chemical anomalies in the ancient Galactic GCs. A more or less extended turnoff region was revealed in many other YMC of the MCs (\cite{milone2009}), and was generally attributed to a stellar age spread.
In 2015 this interpretation suddenly lose its appeal, when researchers considered results of the Geneva tracks and isochrones computed for a wide range  of masses (ages) and rotation rates (e.g. \cite{georgy2013})\footnote{see the page web http://obswww.unige.ch/Recherche/evoldb/index/ created and maintained by C. Georgy and S. Ekstr\"om}. \cite{brandt-huang2015} showed that the turnoff area covered between the non rotating and rotating isochrones increased with the cluster age up to about 1.5\,Gyr, and then decreased, and the same behaviour was displayed by the YMC for increasing ages. In terms of ages this would imply an increasing age spread (\cite{niederhofer2015}), if interpreted with non--rotating models only. 

At the same time, \cite{milone2015} found a different puzzling feature in the color magnitude diagram of the $\sim$400\,Myr cluster NGC\,1856: a ``split" main sequence. This feature could not be understood in terms of differences in age, metallicity or helium content, while splitting and the general features of the color--magnitude diagram were consistent with the superposition of two coeval populations, the first one  including very rapidly rotating stars (the red main sequence and the upper part of the spread turnoff), the other one including slowly rotating stars (the blue main sequence and the lower luminosity turnoff stars) (\cite{dantona2015}). Possibly the slowly rotating stars had also started as rapid rotators, but had been ``braked" by dynamical tides (\cite{zahn1977}) due to interaction with a distant binary companion, as it occurs to the field binaries with orbital periods between 4 and 500\,days (\cite{abtboonyarak2004}). The role of braking was confirmed by \cite{dantona2017}, who remarked that the slowly rotating `blue' MSs in several young clusters required a $\sim$10\% younger age than the rapidly rotating upper `red' turnoff stars. The alternative to a younger population is that the blue MS stars are in a less advanced nuclear burning stage with respect to the requirement of their slow rotation. In other words, these stars follow the evolution of their rapidly rotating counterparts, and feed fresh hydrogen in the H--burning core thanks to rotational mixing, until recent braking leaves them as slow rotators. 
Why the stars in these YMCs are born so rapidly rotating, and what are the precise mechanisms for braking which we see so well in the double MS and in the spread of the turnoff is to be worked out, but it is evident that rotation and its evolution {\sl in the cluster ambient} is an important feature for these stars, and it is described quite well by the Geneva models. 

\cite{milone2018} have shown that the split MS phenomenon occurs for ages from $\sim$30 to  $\sim$400\,Myr, so it involves the range of masses from $\sim$3 to $\sim$9\,M$_\odot$. This range includes the super--AGB and massive AGB masses whose envelopes, processed by Hot Bottom Burning (HBB), are responsible for the formation of second generation stars in the ancient GCs (\cite{ventura2001}), in the so called `AGB scenario' (\cite{dercole2008}), so it is reasonable to assume that rotation, although subject to the quoted braking mechanisms, has been an important ingredient in the evolution of these stars too.
On the other hand, \cite{decressin2009} examined the consequences of rotation on the abundances of intermediate mass AGB and showed that rotational mixing increased so much the abundances of CNO at the second dredge up (2DU) that the AGBs could not be reasonable pollutor sources for the multiple populations. As the AGB scenario is anyway the most adequate to deal with both the formation (\cite{dercole2008}), the chemistry (e.g. \cite{ventura2009}, \cite{ventura2013}, \cite{dantona2016}), the spatial distribution of the different populations (\cite{vesperini2013}), the role of binaries (e.g. \cite{vesperini2011}) ---see also Ventura's review in this book--- in spite of its remaining problems (see, e.g. \cite{renzini2015}), it is now necessary to re-examine the problem of rotating evolution.
\begin{figure}[bt]
\vspace*{-0.5 cm}
\begin{center}
 \includegraphics[width=2.4in]{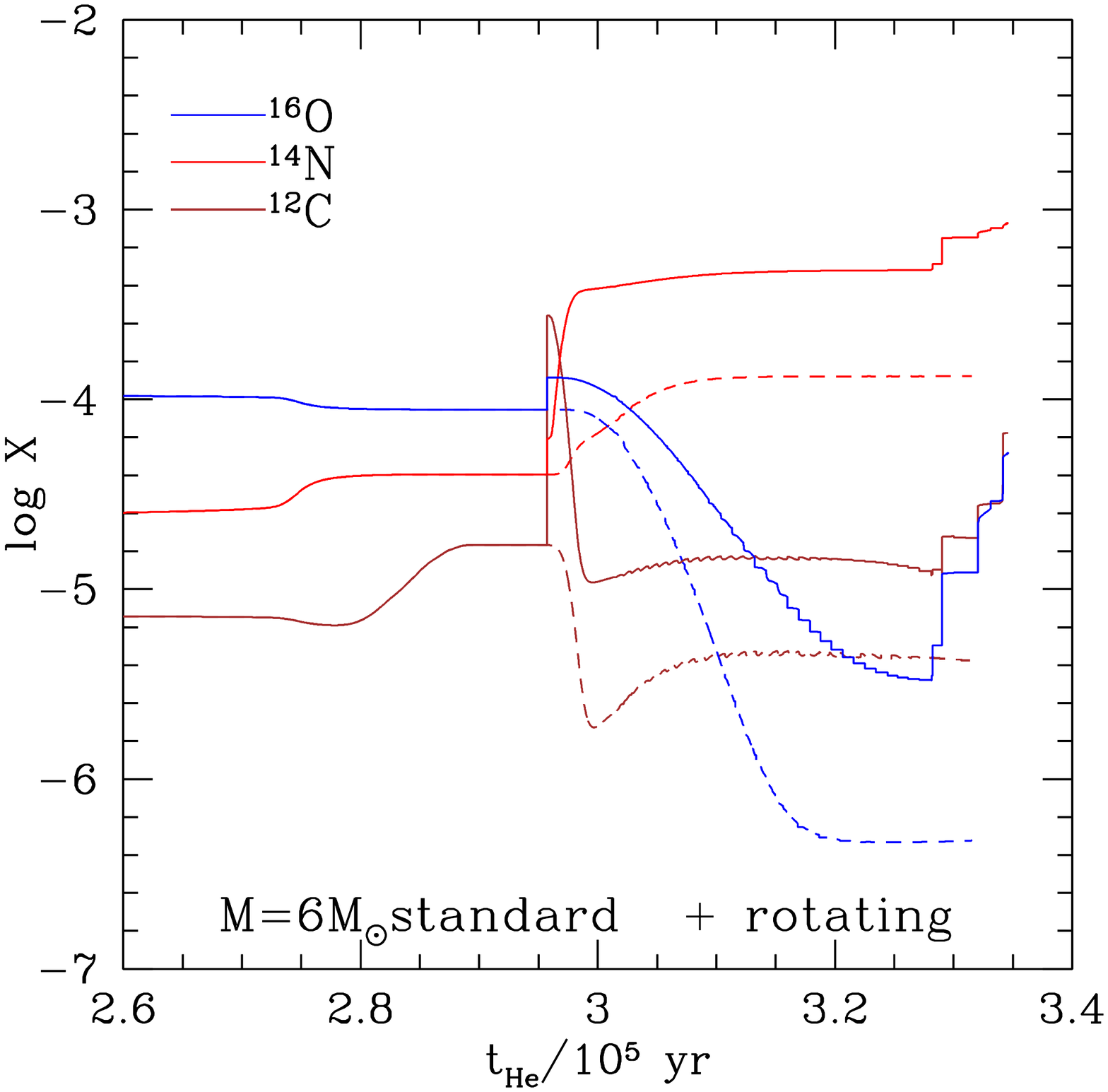} 
 \includegraphics[width=2.4in]{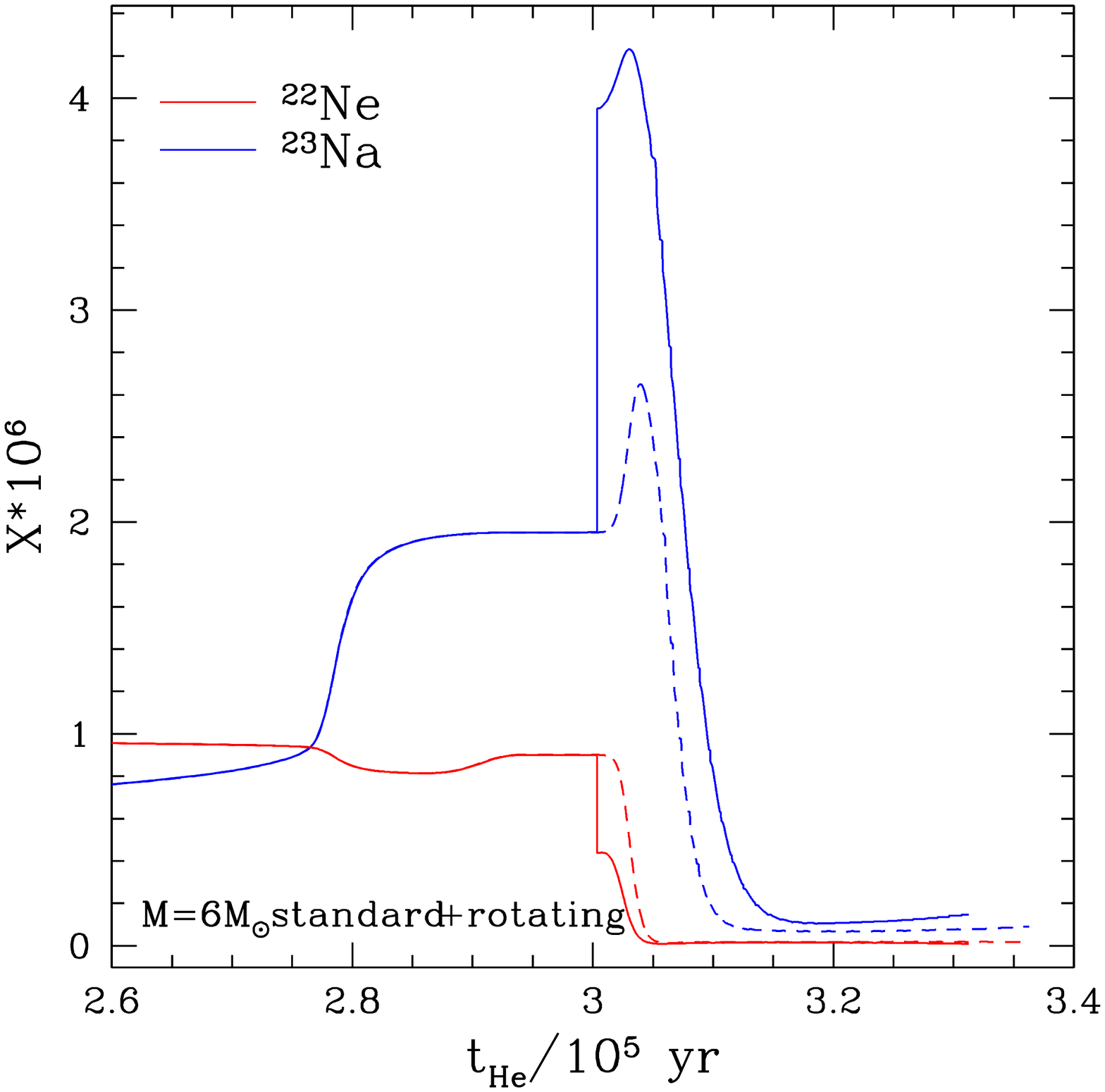} 
\vspace*{-1.35 cm}
 \caption{Evolution of the abundances of C, N and O (left panel) and Na and Ne (right panel) in non rotating model of 6\,M$_\odot$\ (dashed) and when the abundances are increased at the 2DU following rotating MESA models results (full lines). We see that sodium and oxygen survive longer in the rotating case. }
   \label{fig1}
\end{center}
\end{figure}

\begin{figure}[]
\vspace*{-0.5 cm}
\begin{center}
 \includegraphics[width=2.4in]{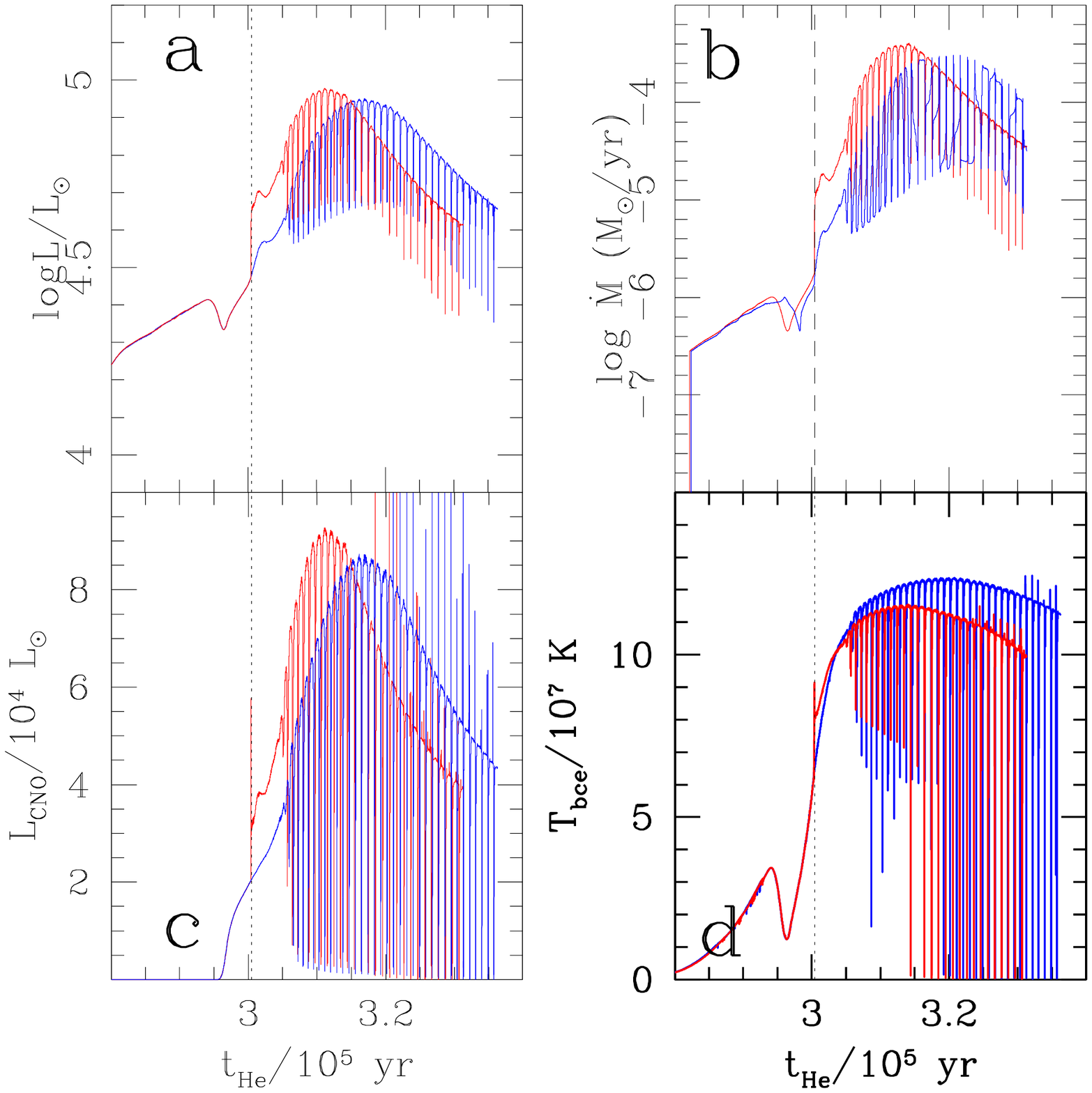} 
 \includegraphics[width=2.4in]{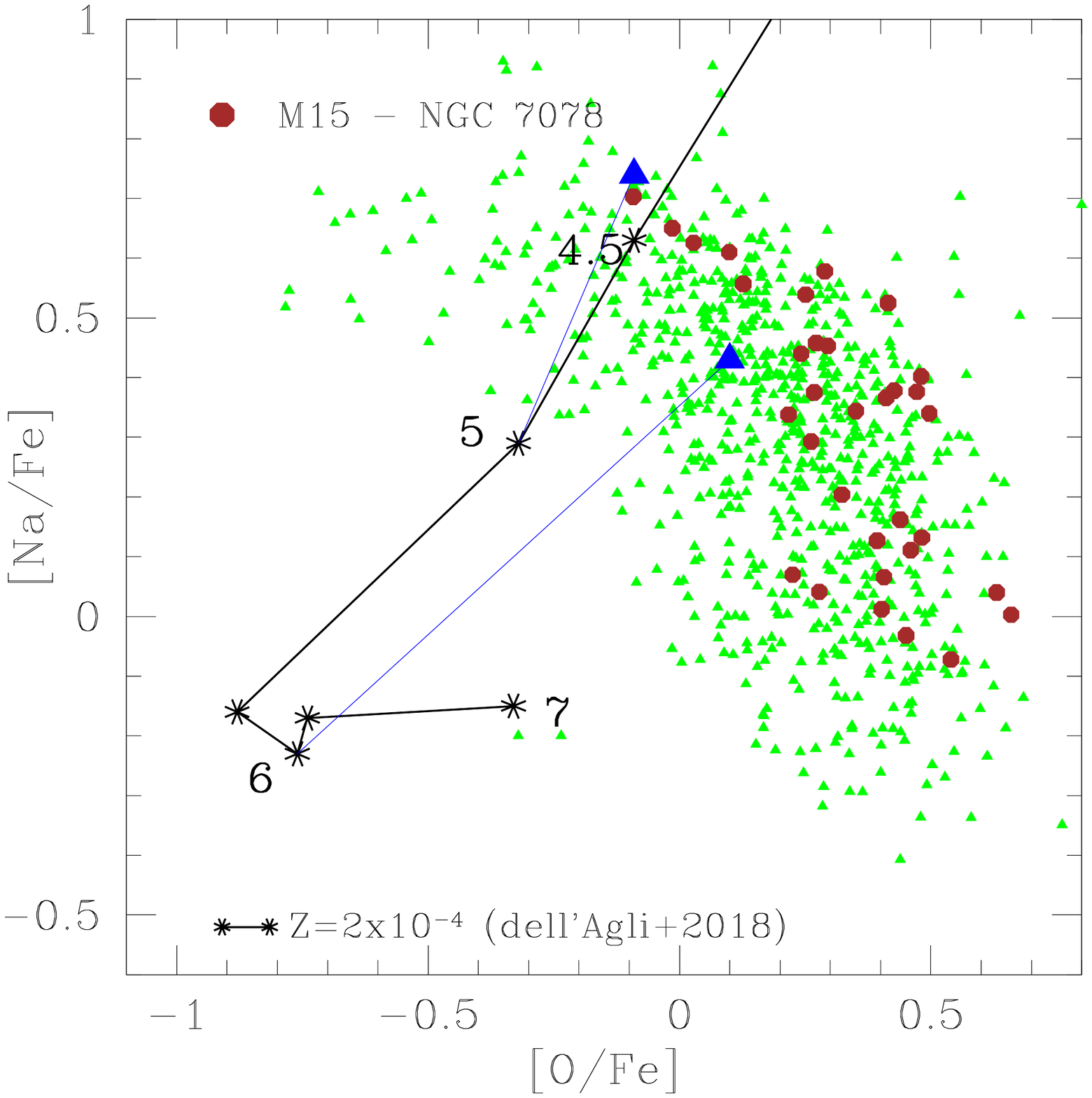} 
\vspace*{-1.35 cm}
 \caption{Left panel: time evolution of  a: log luminosity; b: log mass loss rate; c: CNO luminosity; d: log temperature at the base of the convective envelope. The blue lines refer to the standard track, the red ones refer to the track with increased CNO-Na abundances, simulating the result of rotational mixing. Right panel: in the [Na/Fe] versus [O/Fe], the green triangles refer to all GC data from Carretta et al. 2009, while the brown dots represent the stars in the low emtallicity GC M15. The asterisks connected by black lines are the model yields from Dell'Agli et al. 2018, for the chemistry of M15, and the lines connecting to the blue triangles the yields of the 6 and 5M$_\odot$\ shift in the models with increased CNO-Na abundances.}
   \label{fig1}
\end{center}
\end{figure}

\section{Abundances at the 2DU  in recent rotating models}
While new computations by one of us (Sylvia Ekstr\"om) with the Geneva code more or less reproduced the \cite{decressin2009} results both at very low and intermediate metallicity, the results of the rotating MIST code (\cite{choi2016}),  obtained from the MESA code (\cite{paxton2011}). 
give a much smaller CNO enhancement, it is at most a factor 4--5 for the abundances of the most metal poor GCs ([Fe/H]$\simeq$--2.2). The CNO increase at the metallicities typical of the bulk of the galactic GC population ([Fe/H]$\ge$--1) is negligible, so ---if these models describe the rotational mixing correctly--- we must worry about the difference in the yields only for the lowest metallicity clusters.

As also $^{23}$Na and   $^{22}$Ne abundances are important for the second generation pollution, we considered the whole results in MESA models computed by one of us (Aaron Dotter)  for [Fe/H]=--2 and solar scaled abundances . We used the absolute values of $^{12}$C, $^{14}$N, $^{16}$O, $^{23}$Na and $^{22}$Ne from these models and applied them to the standard (non rotating) models by \cite{dellagli2018} which had the chemical abundances adequate to describe the detailed composition of the GC M\,15 found in the APOGEE survey. By changing the abundances at the 2DU (Fig. 1), we may consider the differential effect of the different chemical abundances on the hot bottom burning phase and on the yields. Obviously, this is not a self--consistent result, but it gives us an interesting way of change in the abundance patterns.

We found out that, for the 6m$_\odot$, the $^{16}$O and the $^{23}$Na are $\sim$0.7\,dex and $\sim$0.6\,dex less depleted than in the standard models, and reproduce better  the abundances in the cluster M\,15, which lacks the extreme O depletions present in intermediate metallicity clusters (Fig. 2, right panel). The 5\,M$_\odot$\, shows a similar effect, by a smaller amount. In fact, the increased CNO mean a larger CNO luminosity at the 2DU (Figure 2, panel c) and thus a larger total L (Panel a), which in turn gives a larger mass loss rate (panel b) during the first evolutionary phases. More mass is then lost when the $^{23}$Na is high, and its yield is larger. Later on, however, the larger mass loss rate provide a smaller temperature at the bottom of convection (panel d) , and therefore a slower conversion of oxygen to nitrogen. Obviously, these results depend very much on the precise CNO and Na increase at the 2DU, and  the total CNO increase assumed here are too large to be consistent with the CNO abundance determinations in the cluster, but the trends we find in this preliminary models deserve further investigation.




\end{document}